\documentclass[runningheads,a4paper]{llncs}

\pdfoutput=1

\usepackage{amsmath,amssymb}

\usepackage[english]{babel}
\usepackage{graphicx}
\usepackage{url}

\begin{document}

\mainmatter

\title{Towards Crowdsourcing and Cooperation in Linguistic Resources}

\author{Dmitry Ustalov\inst{1,2,3}}

\institute{
Krasovsky Institute of Mathematics and Mechanics, Ekaterinburg, Russia,
\and
Ural Federal University, Ekaterinburg, Russia,
\and
NLPub, Ekaterinburg, Russia\\
\email{dau@imm.uran.ru}
}

\maketitle

\begin{abstract}
Linguistic resources can be populated with data through the use of such approaches as crowdsourcing and gamification when motivated people are involved. However, current crowdsourcing genre taxonomies lack the concept of cooperation, which is the principal element of modern video games and may potentially drive the annotators' interest. This survey on crowdsourcing taxonomies and cooperation in linguistic resources provides recommendations on using cooperation in existent genres of crowdsourcing and an evidence of the efficiency of cooperation using a popular Russian linguistic resource created through crowdsourcing as an example.

\end{abstract}

\section{Introduction}

Crowdsourcing has become a mainstream and well-suited approach for solving many linguistic data gathering problems such as sense inventory creation \cite{Biemann:13}, corpus annotation \cite{Sabou:14}, information extraction \cite{Lofi:12}, etc. However, its most effective use still remains a problem because human annotators' motivation and availability are tantalizingly constrained and it is crucial to get the most of performance from the effort interested people can make.

Another extremely popular term nowadays is \textit{gamification}. The origin of the gamification concept is, of course, video game industry. The idea of gamification is in embedding interactive and game-based techniques into application to strengthen user engagement and increase the time spent annotating. Due to the insufficiency of exploration, gamification is more rarely used in academia when compared to the industry.

Cooperation is a major, if not principal, element of today's video games, which is confirmed by the presentations made in recent years at \textsc{E3} --- the largest video game exposition and event. Initially, multiplayer mode in video games was focused on \textit{player versus player} competitions, but a few years ago the focus has changed to \textit{cooperated human players versus AI} and \textit{guild versus guild} games.

The work, as described in this paper, makes the following contributions: 1) it presents a survey on crowdsourcing taxonomies and cooperation in linguistic resources, 2) makes recommendations on using cooperation in existent genres of crowdsourcing, and 3) provides an evidence of the efficiency of cooperation represented by a popular Russian linguistic resource created through crowdsourcing.

The rest of this paper is organized as follows. Section \ref{sec:genres} focuses on related work towards crowdsourcing genres and cooperation in linguistic resources. Section \ref{sec:cooperation} is devoted to the cooperative aspect of crowdsourcing. Section \ref{sec:evidence} discusses cooperation using OpenCorpora as the example, which is a sufficiently popular Russian linguistic resource created through crowdsourcing. Section \ref{sec:discussion} interprets and explains the obtained results. Section \ref{sec:conclusion} concludes with final remarks and directions for the future work.

\section{Crowdsourcing Genres \& Activities}
\label{sec:genres}

Early studies on crowdsourcing genres in their wide definition were conducted in 2009. Quinn \& Bederson in their technical report \cite{Quinn:09} proposed the term of \textit{distributed human computation} along with the taxonomy of seven different genres of these computations such as games with a purpose, mechanized labor, wisdom of crowds, crowdsourcing, dual-purpose work, grand search, human-based genetic algorithms, and knowledge collection from volunteer contributions.

In the same year Yuen et al. also presented \cite{Yuen:09} another taxonomy of five crowdsourcing genres: initiatory human computation, distributed human computation, social game-based human computation with volunteers, paid engineers and online players, which is similar to the previously mentioned.

Many studies following the early ones are focused on classification of whether a crowdsourced project belongs to a specific class of the given taxonomy. For instance, Sabou et al. study of correlation between crowdsourcing genres \cite{Sabou:12}, quality assessment \cite{Sabou:13}, and guidelines on corpus annotation through crowdsourcing \cite{Sabou:14} align various best practices among the established genres.

There are other attempts to create a taxonomy of crowdsourcing genres. Zwass investigated the phenomena of \textit{co-creation} \cite{Zwass:10} and proposed a taxonomy of user-created digital content which includes the following: knowledge compendia, consumer reviews, multimedia content, blogs, mashups, virtual worlds. The resulted taxonomy appears to be too general and, since it was not intended, does not fit the natural language processing field perfectly.

Erickson presented \textit{four quadrant model} \cite{Erickson:11} composed of two orthogonal dichotomies to classify crowdsourcing projects: ``same place--different places'' and ``same time--different times''. The resulted taxonomy tends to assign all the mentioned above crowdsourced projects to the ``different places--different times'' quadrant also called \textit{Global Crowdsourcing}.

Some studies propose much narrower dichotomies. This is the case of the research conducted by Suendermann \& Pieraccini \cite{Suendermann:13}, which introduces a concept of \textit{private crowd} being a trade-off between two extremes: an inexpensive, highly available yet uncontrolled \textit{public crowd} such as the Amazon's one, and the expensive to hire, high-quality and professional expert annotators. The \textit{private crowd} term can be referred to as \textit{controlled crowd}.

\subsection{Three Genres of Crowdsourcing}

In 2013, Wang et al. aggregated most of the previous studies in their very well-done survey. The mentioned work emphasizes three intuitive and well-separated genres of crowdsourcing \cite{Wang:13}:

\begin{description}
  \item[Games with a purpose] (\textsc{GWAPs}), when a player without any special knowledge is put into a gaming environment and have to make right decisions to win the game under the pressure of time or any game mechanics' constraints. Phrase Detectives\footnote{\url{https://anawiki.essex.ac.uk/phrasedetectives/}} and JeuxDeMots\footnote{\url{http://www.jeuxdemots.org/}} can be considered as good examples of such games.
  \item[Mechanized labor] (\textsc{MLab}), when an annotator who meet the preliminary requirements is asked to answer a questionnaire on a centralized platform and is rewarded for their work by micropayments. The most well-recognized examples of \textsc{MLab} are Amazon Mechanical Turk\footnote{\url{http://mturk.com/}} and CrowdFlower\footnote{\url{http://www.crowdflower.com/}}.
  \item[Wisdom of the crowd] (\textsc{WotC}), when motivated volunteers share their knowledge on the given topic in the free form in order to answer some question, to explain something to other people, and so on. The obvious examples of \textsc{WotC} are Wikipedia\footnote{\url{http://wikipedia.org/}} and Yahoo! Answers\footnote{\url{https://answers.yahoo.com/}}.
\end{description}

Observations reveal that research papers often do not specify the exact crowdsourcing genre and treat the crowdsourcing term as a synonym to \textsc{MLab} due to extreme popularity of the Amazon's product.

\subsection{Cooperation in Linguistic Resources}

Cooperation, derived from \textit{to cooperate}, is to work actively with rather than against others \cite[p. 435]{Alfie:92}. Unfortunately, cooperative crowdsourcing in linguistic resources is less explored in the literature, but present studies show that considering the concept of cooperation in crowdsourcing is a trending topic deserving attention.

An early study of Wikipedia and its quality by Wilkinson \& Huberman \cite{Wilkinson:07} found a statistically significant correlation between page edits, talkpage conversations and the quality of these pages. The study revealed the fact that pages with more intense discussion activity often have better quality than less discussed ones.

A study by Arazy \& Nov \cite{Arazy:10} pays a special attention to \textit{local inequality} --- inequality of editors' contribution in a particular article, and \textit{global inequality} --- inequality in overall Wikipedia activity for the same set of editors. As a result, they found that global inequality has an impact on local inequality, which influences editors' coordination in a positive way, which in its turn contributes to quality.

Budzise-Weaver et al. \cite{BudziseWeaver:12} consider several cases of multilingual digital libraries and their collaboration both with state institutions and crowdsourced projects in order to provide multilingual information access for users. The paper does not describe how exactly crowdsourcing can help digital libraries in doing their job, but does demonstrate significant interest to crowdsourcing from an interdisciplinary point of view.

Ranj Bar \& Maheswaran \cite{RanjBar:14} in their case study on Wikipedia concluded that new mechanisms are needed to coordinate the activities in crowdsourcing due to the fact that high quality articles are controlled by small groups of permanent editors, and supporting these articles is a huge burden for the editors.

\section{Crowdsourcing Genres and Cooperation}
\label{sec:cooperation}

Each of the three crowdsourcing genres has its own identities; and the principle of paritipants' cooperation changes with each particular crowdsourcing instance. However, it seems possible to denote three common points:

\begin{itemize}
  \item \textbf{attractiveness}, the degree of how a participant can find a crowdsourcing process attractive,
  \item \textbf{usefulness}, the degree of how a participant can find his activity results useful to their own purposes,
  \item \textbf{difficulty}, the degree of how it is difficult to embed cooperative elements into a process.
\end{itemize}

When specific case studies are available, the correspondent details are provided.

\subsection{Games with a purpose}

The main advantage of \textsc{GWAPs} is their \textit{high attractiveness}, because people love video games and it is easier to get new participants than in other genres of crowdsourcing. One may find \textit{low usefulness} in these games, but the more attractive the game is, the less other factors are becoming important.

It is necessary to mention that video games are a very costly kind of software and producing \textsc{GWAPs} requires not only creating a game, but also designing innovative game mechanics allowing a player to both enjoy the game and to implicitly produce valuable data. Thus, games with a purpose have \textit{high difficulty} to be realized.

Authors of Phrase Detectives say that the cost of data gathering using their means is lower than using other approaches \cite{Poesio:13}, but they did not consider the total cost of the game design and development. Elements of real-time players' cooperation may enhance \textsc{GWAPs} attractiveness even more. The evidence of this is the fact that modern cooperative multiplayer video games like Dota~2 or Destiny have substituted traditional \textit{free for all} (deathmatch) multiplayer games.

\subsection{Mechanized labor}

Since \textsc{MLab} projects are often deployed on specialized platforms available on the World Wide Web, the main advantage of \textsc{MLab} is its \textit{low difficulty}: cooperative elements may be embedded supplementarily to the annotation process through allowing annotators to join teams and making them participate in the team-based activity.

In order to cover as much domains as possible, platforms' owners provide only very utilitarian and generic interfaces allowing one to answer a questionnaire without exposing them to any domain-specific features.

Since \textsc{MLab} participants are often rewarded for their work that may be or may not be interesting for them, the mechanized labor projects have \textit{medium usefulness} and usually \textit{low attractiveness}.

\subsection{Wisdom of the crowd}

The strong side of \textsc{WotC} projects is, indeed, \textit{high usefulness} due to the fundamental principle of such a genre, when volunteers make efforts to make their resource better for everyone. \textsc{WotC} have \textit{low attractiveness} for the same reasons, however it depends on every particular instance.

The above mentioned study by Arazy \& Nov also touches upon a typical regulation problem called ``edit warring'' in Wikipedia \cite{Arazy:10}, when ``editors who disagree about the content of a page repeatedly override each other's contributions, rather than trying to resolve the disagreement through discussion''.

The phenomena of ``edit warring'' was later studied by Yasseri et. al \cite{Yasseri:12}. Such a problem may be partially resolved by using the controlled crowd instead of the public one when volunteers have a mentor and responsibility for their actions \cite{Braslavski:14}. Therefore, such projects have \textit{medium difficulty}.

\section{Evidence}
\label{sec:evidence}

An evidence that cooperation does work and really stimulates participants to do more assignments is the case of OpenCorpora, which is a project focused on creation of a large annotated Russian corpus through crowdsourcing \cite{Bocharov:13}.

Currently, OpenCorpora participants have to annotate morphologically ambiguous examples in the \textsc{MLab} manner. One can annotate examples individually, but has an opportunity to join teams and annotate examples in cooperation with their teammates. A team can be created and joined by everyone, and teams challenge each other by means of active collaborators, annotated examples, and error rates.

As according to the full-scale pilot study conducted on one of the largest Russian information technologies' websites\footnote{\url{http://habrahabr.ru/post/152799/#comment_5315923}}, volunteers were very positive about their participation in the cooperative annotation. The study was followed by the creation of the largest team uniting $170$ participants. The team got the 2\textsuperscript{nd} place in the total rank\footnote{\url{http://opencorpora.org/?page=stats}} based on the number of the annotated examples.

The possible explanation of such a result would be found in what have driven the participants' motivation. It was not only their altruism and readiness to help, but the possibility for their team to get the leading places in the total rank, as well as their personal participation being one of the keys to the team's possible success.

\begin{figure}[t]
  \centering
  \includegraphics[width=\textwidth]{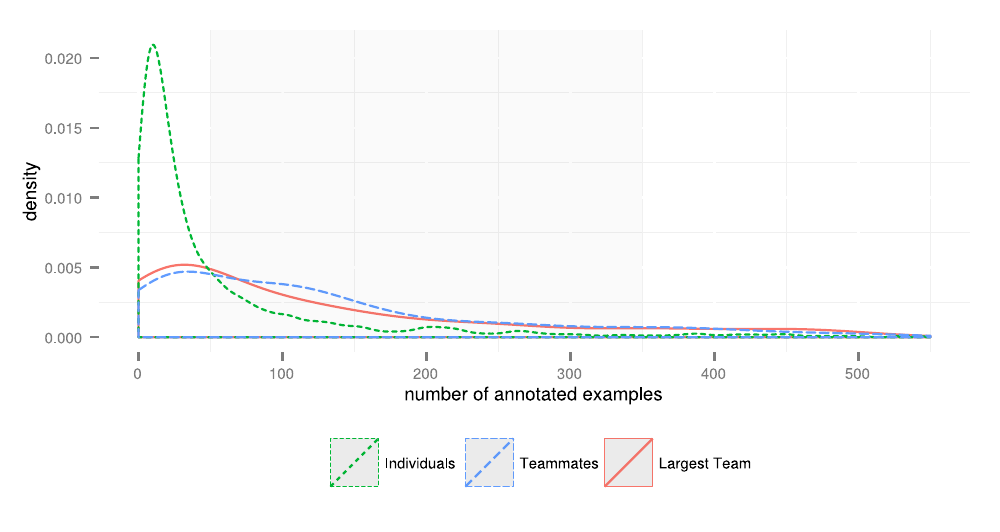}
  \caption{Annotated examples' densities by three user groups: the individuals, the teammates, and the largest team; the highlighted region corresponds to the interval between $50$ and $350$ examples.}
  \label{fig:users}
\end{figure}

\subsection{``Is there a relationship?''}

To make it possible to study the present result more thoroughly, the OpenCorpora team has kindly provided us with the dataset consisted of user ID, the group's name, and various activity information including total number of the annotated examples per user. Hereafter participants who joined a team are referred to as \textit{teammates}, and those who did not join a team are referred to as \textit{individuals}.

The initial dataset contains information on $2642$ users: $2219$ of them are individuals and $423$ are teammates. The distributions' densities are depicted at Fig.~\ref{fig:users} and seem to be right-skewed. In order to remove outliers from the dataset, users who annotated less than $50$ examples or more than $350$ examples have been excluded. As a result, the dataset has been reduced to $579$ individuals and $195$ teammates, $71$ of those are the members of the largest team.

In general, the individuals annotated $801\,531$ examples and the teammates annotated $970\,650$, while in the dataset the individuals annotated $71\,150$ examples and the teammates annotated only $29\,049$ examples.

Hence, the research question is \textit{``Is there a relationship between being a team member and the number of annotated examples for a regular OpenCorpora user?''}

\subsection{Inference}

\begin{figure}[t]
  \centering
  \includegraphics[width=\textwidth]{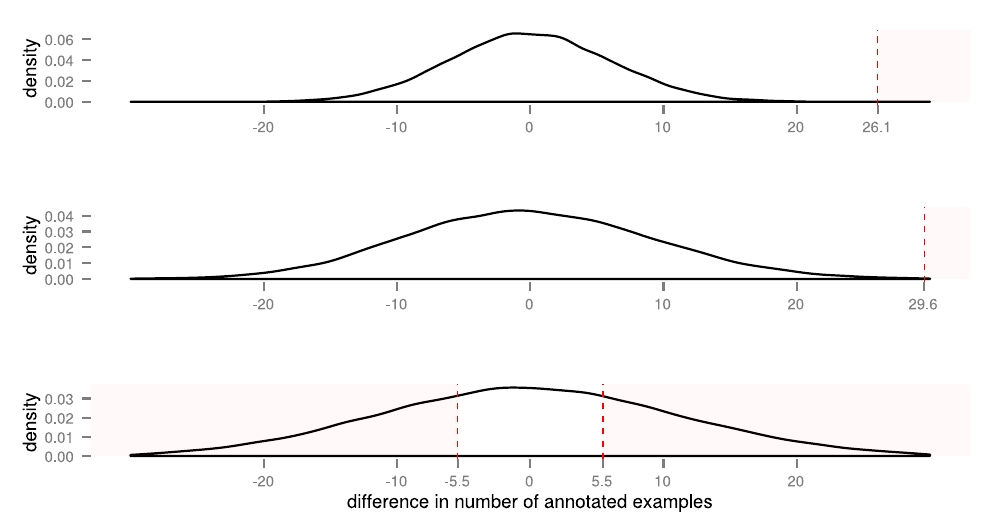}
  \caption{Simulated differences in number of annotated examples, the vertical dashed lines represent observed differences: the upper plot (a) corresponds to $H$, the middle plot (b) corresponds to $H'$, and the lower (c) corresponds to $H''$.}
  \label{fig:hypotheses}
\end{figure}

Since the dataset is right-skewed and such hypothesis tests as \textit{t-test} may be unreliable, a randomization test was implemented in the R programming language and executed for $25\,000$ times under the significance level of $\alpha = .05$ in order to estimate the unbiased \textit{p-value}.

The true difference in means of the numbers of annotated examples among the teammates ($\mu_T$) and the individuals ($\mu_I$) has been examined. The following hypothesis $H$ was evaluated in order to find a relationship between being a team member and the number of annotated examples:

\begin{description}
  \item[$H_0:$] $\mu_T - \mu_I = 0$, the teammates and the individuals on average have no difference in their annotation activity,
  \item[$H_A:$] $\mu_T - \mu_I > 0$, the teammates tend to annotate more examples on average than the individuals.
\end{description}

The density of differences in the number of annotated examples is demonstrated at Fig~\ref{fig:hypotheses}(a): the observed difference in means for this one-tailed test is $\bar{x}_T - \bar{x}_I = 26.085$, and the \textit{p-value} is $p = 0$. Thus, $p < \alpha$ and the null hypothesis $H_0$ has been rejected, suggesting that $\mu_T > \mu_I$: the teammates tend to annotate more examples than the individuals.

\section{Discussion}
\label{sec:discussion}

The obtained result can also be explained by a teammate being more loyal and attached to the resource than an individual. Therefore, it is reasonable to study the performance of a particular team.

\subsection{The Largest Team vs. The Individuals}

In order to compare the behavior of the individuals and the largest team members instead of all the teammates, the true difference in means of the numbers of annotated examples among the teammates of the largest team ($\mu_H$) and the individuals ($\mu_I$) was examined. The following hypothesis $H'$ was evaluated in the similar way as the previous one:

\begin{description}
  \item[$H'_0:$] $\mu_H - \mu_I = 0$, the teammates of the largest team and the individuals on average have no difference in their annotation activity,
  \item[$H'_A:$] $\mu_H - \mu_I > 0$, the teammates of the largest team annotate more examples on average than the individuals.
\end{description}

The simulation results for this one-tailed test are presented at Fig.~\ref{fig:hypotheses}(b): the observed difference in means is $\bar{x}_H - \bar{x}_I = 29.552$ and the \textit{p-value} is $p = .001$. Since $p < \alpha$, the null hypothesis $H'_0$ has been rejected, suggesting that $\mu_H > \mu_I$: the teammates of the largest team annotate more examples than the individuals.

This result agrees well with the $H_0$ hypothesis and can be explained by the fact that the largest team is still relatively small and consists of only $170$ teammates who were highly motivated for a short time due to news rotation on the website where they came from. Their activity decreased significantly when the announcement of the OpenCorpora disappeared from the news headline. Their team took the 2\textsuperscript{nd} place on the leaderboard; they lost to the the leading team as the latter had annotated approximately seven times more examples ($501\,963$ versus $76\,559$).

\subsection{The Largest Team vs. Other Teams}

Statistical testing of teams' performance based on comparison of their impact is complicated due to lack of participants in other teams. For instance, the 2\textsuperscript{nd} largest team is comprised of $36$ users only, the 3\textsuperscript{rd} largest --- $24$, the 4\textsuperscript{th} --- $13$, which is insufficient for any meaningful test. However, it is indeed possible to compare the performance of the largest team with the performance of other teams considered together.

The true difference in means of the numbers of annotated examples among the teammates of the largest team ($\mu_H$) and other teams ($\mu_R$) was examined, and the following hypothesis $H''$ has been evaluated:

\begin{description}
  \item[$H''_0:$] $\mu_H - \mu_R = 0$, the teammates of the largest team and other teammates on average have no difference in their annotation activity,
  \item[$H''_A:$] $\mu_H - \mu_R \neq 0$, the teammates of the largest team and other teammates on average have the difference in their annotation activity.
\end{description}

The simulation results for this two-tailed test are presented at Fig.~\ref{fig:hypotheses}(c): the observed difference in means is $\bar{x}_H - \bar{x}_R = 5.453$ and the \textit{p-value} is $p = .629$. Since $p > \alpha$, the null hypothesis $H'_0$ has not been rejected, suggesting that $\mu_H = \mu_R$: the teammates of the largest team annotate the same number of examples as other teammates do.

\section{Conclusion}
\label{sec:conclusion}

According to the obtained results, there \textit{is} a correlation between being a team member and the number of annotated examples for a regular OpenCorpora user. The use of team-based cooperation can improve the user activity on crowdsourced linguistic resources. However, since the study is observational, it was impossible to establish causal relationships between the variables.

When organized in teams, users do provide more annotations comparing with those who are not organized in teams. Thus, it is highly recommended for a crowdsourced resource to provide users with the opportunity to join teams and annotate examples in cooperation with their teammates.

The statistical hypotheses have been evaluated with use of the randomization test with the significance level of $.05$. The present dataset is available\footnote{\url{http://ustalov.imm.uran.ru/pub/opencorpora-cooperation.tar.gz}} in an depersonalized form under the Creative Commons Attribution-ShareAlike 3.0 license. The source code of the above mentioned simulation program is included under the MIT License.

Further work may be focused on assessing the quality of team-based cooperation results and on studying the patterns of cooperation and the efficiency of their use in other popular linguistic resources created through crowdsourcing.

\subsubsection*{Acknowledgments.}\sloppy
This work is supported by the Russian Foundation for the Humanities, project no.~13-04-12020 ``New Open Electronic Thesaurus for Russian'', and by the Program of Government of the Russian Federation 02.A03.21.0006 on 27.08.2013.

\pagebreak

The author would like to thank Dmitry Granovsky for the extended statistical information collected from \url{http://opencorpora.org/}. The author is also grateful to the anonymous referees who offered very useful comments on the present paper.

\bibliographystyle{splncs}
\bibliography{ustalov.russir2014}

\end{document}